# Dark Matter and Cosmology: CDM with a Cosmological Constant (ΛCDM) vs. CDM with Hot Dark Matter (CHDM) [*]

Joel Primack[a] and Anatoly Klypin[b]

[a]Physics Department, University of California, Santa Cruz,CA 95064; joel@physics.ucsc.edu

[b]Department of Astronomy, New Mexico State University, Las Cruces, NM 88001; aklypin@nmsu.edu

The fact that the simplest modern cosmological theory, standard Cold Dark Matter (sCDM), almost fits all available data has encouraged the search for variants of CDM that can do better. Here we discuss what are perhaps the two most popular variants of CDM that might agree with the data: ΛCDM and CHDM. While the predictions of COBE-normalized ΛCDM and CHDM both agree well with the available data on scales of ∼ 10 to 100 $h^{-1}$ Mpc, each has potential virtues and defects. ΛCDM with $\Omega_0 \sim 0.3$ has the possible virtue of allowing a higher expansion rate $H_0$ for a given cosmic age $t_0$, but the defect of predicting too much fluctuation power on small scales. CHDM has less power on small scales, so its predictions appear to be in good agreement with data on the galaxy distribution, but it remains to be seen whether it predicts early enough galaxy formation to be compatible with the latest high-redshift data. Also, several sorts of data suggest that neutrinos have nonzero mass. And two very recent observational results favor high cosmic density, and thus favor $\Omega = 1$ models such as CHDM over ΛCDM — (1) the positive deceleration parameter $q_0 > 0$ measured using high-redshift Type Ia supernovae, and (2) the low primordial deuterium/hydrogen ratio measured in two different quasar absorption spectra. If confirmed, (1) rules out a cosmological constant large enough to help significantly with the $H_0$-$t_0$ problem; while (2) suggests that the baryonic cosmological density is at the upper end of the range allowed by Big Bang Nucleosynthesis, perhaps high enough to resolve the "cluster baryon crisis" for $\Omega = 1$ models. We try to identify "best" variants of both ΛCDM and CHDM, and discuss critical observational tests for both models.

## 1. INTRODUCTION AND SUMMARY

"Standard" $\Omega = 1$ Cold Dark Matter (sCDM) with a near-Zel'dovich spectrum of primordial fluctuations [1] seemed to many theorists to be the most attractive of all modern cosmological models. But while sCDM normalized to the amplitude of cosmic microwave background fluctuations observed by COBE is a good fit to most large-scale observations, it is inconsistent with data on smaller scales [2]. Here we discuss what are perhaps the two most popular variants of sCDM that might agree with the data: ΛCDM and CHDM. The linear *matter* power spectra for these two models are compared in Figure 1 with the real-space *galaxy* power spectrum obtained from the two-dimensional APM galaxy power spectrum [3]. The ΛCDM and CHDM models essentially bracket the range of power spectra in currently popular cosmological models which are variants of CDM.

ΛCDM cosmological models with a positive cosmological constant ($\Lambda > 0$) and $\Omega_0 = 1 - \Omega_\Lambda \approx 0.3$, where $\Omega_\Lambda \equiv \Lambda/(3H_0^2)$, have been advocated [4] because they allow a larger Hubble constant $H_0$ for a given age of the universe $t_0$, and they predict a larger fraction of baryons in galaxy clusters, than $\Omega = 1$ models. Early galaxy formation also is considered to be a desirable feature of these models. But early galaxy formation implies that fluctuations on scales of a few Megaparsecs spent more time in the nonlinear regime, as compared with Cold + Hot Dark Matter (CHDM) models. As has been known for a long time, this results in excessive clustering on small scales. We have found that a typical ΛCDM model with $H_0 = 70$ km s$^{-1}$ Mpc$^{-1}$, $\Omega_0 = 0.3$, and cosmological constant $\Lambda$ such that $\Omega_\Lambda \equiv \Lambda/(3H_0^2) = 1 - \Omega_0$, normalized to COBE on large scales (this fixes $\sigma_8 \approx 1.1$ for this model), is compatible with the number-density of galaxy clusters, but predicts a power spectrum of galaxy clustering in real space which is much too high for wavenumbers $k = (0.4 - 1)h/\text{Mpc}$ [5]. This conclusion holds if we assume either that galaxies trace the dark matter, or just that a region with higher density produces more galaxies than a region with lower density. One can see immediately from Figure 1 that there will be a problem with this ΛCDM model, since the APM power spectrum is approximately equal to the linear power spectrum at wavenumber $k \approx 0.6h \text{ Mpc}^{-1}$, so there is no room for the extra power that nonlinear evolution will certainly produce on this scale (see Figure 1 of Ref. [5] and further discussion below). The only way to reconcile the model with the observed power spectrum is to assume that some mechanism causes strong anti-biasing — i.e., that regions with high dark matter density produce fewer galaxies than regions with low density. While theoretically possible, this seems very unlikely; biasing rather than anti-biasing is expected on small scales [6].

---

[*]To be published in the *Proceedings of the International Conference on Sources and Detection of Dark Matter in the Universe*, UCLA, February, 1996, D. Cline and D. Sanders, editors, Nucl. Phys. B, Proceedings Supplement, 1996.

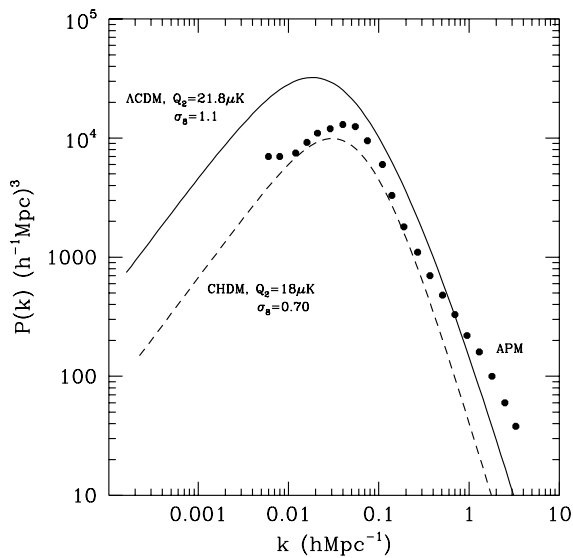

Figure 1. Power spectrum of dark matter for ΛCDM and CHDM models considered in this paper, both normalized to COBE, compared to the APM galaxy real-space power spectrum.

Our motivation to investigate this particular ΛCDM model was to have $H_0$ as large as might possibly be allowed in this class of models, which in turn forces $\Omega_0$ to be rather small in order to have $t_0 \gtrsim 13$ Gyr. There is little room to lower the normalization of this ΛCDM model by tilting the primordial power spectrum $P_p(k) = Ak^{n_p}$, i.e., assuming $n_p$ significantly smaller than the "Zel'dovich" value $n_p = 1$, since then the fit to data on intermediate scales will be unacceptable — e.g., the number density of clusters will be too small [5]. Tilted ΛCDM models with higher $\Omega_0$, and therefore lower $H_0$ for $t_0 \lesssim 13$ Gyr, appear to have a better hope of fitting the available data, based on comparing quasi-linear calculations to the data [5,7]. But all cosmological models with a cosmological constant Λ large enough to help significantly with the $H_0$-age problem are in trouble with observations providing strong upper limits on Λ [9]: gravitational lensing [8], number counts of elliptical galaxies [10], and especially the preliminary results from measurements of the cosmological deceleration parameter $q_0$ using high-redshift Type Ia supernovae [11].

CHDM cosmological models with $\Omega = 1$ mostly in cold dark matter but with a small admixture of hot dark matter (light neutrinos with a total mass of about 5 eV contributing $\Omega_\nu \approx 0.2$ for $h = 0.5$, where $h \equiv H_0/(100 \text{ km s}^{-1} \text{ Mpc}^{-1})$) are a good fit to much observational data [12,13] — for example, correlations of galaxies and clusters and direct measurements of the power spectrum $P(k)$, velocities on small and large scales, and other statistics such as the Void Probability Function (probability $P_0(r)$ of finding no bright galaxy in a randomly placed sphere of radius $r$). We had earlier shown that CHDM with $\Omega_\nu = 0.3$ predicts a VPF larger than observations indicate [14], but new results based on our $\Omega_\nu = 0.2$ simulations in which the neutrino mass is shared equally between two neutrino species [12] show that the VPF for this model is in excellent agreement with observations. However, our simulations [5] of COBE-normalized ΛCDM with $H_0 = 70$ and $\Omega_0 = 0.3$ lead to a VPS which is not compatible with the data [15].

Moreover, there is mounting astrophysical and laboratory data suggesting that neutrinos have non-zero mass [12,16]. The analysis of the LSND data through 1995 [17] strengthens the signal for $\bar{\nu}_\mu \to \bar{\nu}_e$; comparison with exclusion plots from other experiments implies a lower limit $\Delta m^2_{\mu e} \equiv |m(\nu_\mu)^2 - m(\nu_e)^2| \gtrsim 0.2$ eV$^2$, implying in turn a lower limit $m_\nu \gtrsim 0.45$ eV, or $\Omega_\nu \gtrsim 0.02(0.5/h)^2$. This implies that the contribution of hot dark matter to the cosmological density is larger than that of all the visible stars ($\Omega_* \approx 0.004$ [18]). The Kamiokande data showing that the deficit of atmospheric muon neutrinos increases with zenith angle suggests that $\nu_\mu \to \nu_\tau$ oscillations occur with an oscillation length comparable to the height of the atmosphere, implying that $\Delta m^2_{\tau\mu} \sim 10^{-2}$ eV$^2$ [19] — i.e., that $\nu_\mu$ and $\nu_\tau$ are nearly degenerate in mass, consistent with the hot dark matter mass being shared between two neutrino species. Because free streaming of the neutrinos damps small-scale fluctuations, even a little hot dark matter causes reduced fluctuation power on small scales and requires substantial cold dark matter to compensate; thus evidence for even 2 eV of neutrino mass favors large $\Omega$ and would be incompatible with $\Omega_c$ as small as 0.3 [12]. Allowing $\Omega_\nu$ and the tilt to vary, CHDM can fit observations over a somewhat wider range of values of the Hubble parameter $h$ than standard or tilted CDM [20], especially if the neutrino mass is shared between two or three neutrino species [12,21]. However, as for all $\Omega = 1$ models, $h$ larger than about 0.5 conflicts with $t_0 \gtrsim 13$ Gyr from globular cluster and white dwarf cooling age estimates. Another consequence of the reduced power on small scales is that structure formation is more recent in CHDM than in $\Lambda$CDM. This may conflict with observations of protogalaxies at high redshift, although the available evidence does not yet permit a clear decision on this (see below). The evidence from preliminary data of a fall-off of the amount of neutral hydrogen in damped Lyman $\alpha$ systems for $z \gtrsim 3$ [22] is in accord with predictions of CHDM [23].

## 2. CLUSTER BARYONS

Primack has recently reviewed the astrophysical data bearing on the values of the fundamental cosmological parameters, especially $\Omega_0$ [9]. One of the arguments against $\Omega = 1$ which seemed hardest to answer was the "cluster baryon crisis" [24]: recent X-ray observations of clusters show that the abundance of baryons, mostly in the form of gas (which typically amounts to several times the total mass of the cluster galaxies), is as much as 20% if $h$ is as low as 0.5. For the Coma cluster the baryon fraction within the Abell radius is

$$f_b \equiv \frac{M_b}{M_{tot}} \geq 0.009 + 0.050 h^{-3/2}, \qquad (1)$$

where the first term comes from the galaxies and the second from gas. If clusters are a fair sample of both baryons and dark matter, as they are expected to be based on simulations, then this is 2-3 times the amount of baryonic mass expected on the basis of BBN in an $\Omega = 1$, $h \approx 0.5$ universe, though it is just what one would expect in a universe with $\Omega_0 \approx 0.3$. The fair sample hypothesis implies that

$$\Omega_0 = \frac{\Omega_b}{f_b} = 0.33 \left(\frac{\Omega_b}{0.05}\right)\left(\frac{0.15}{f_b}\right). \qquad (2)$$

A recent review of gas in a sample of clusters [25] finds that the baryon mass fraction within about 1 Mpc lies between 10 and 22% (for $h = 0.5$; the limits scale as $h^{-3/2}$), and argues that it is unlikely that (a) the gas could be clumped enough to lead to significant overestimates of the total gas mass — the main escape route considered in [24] (cf. also [26]). If $\Omega = 1$, the alternatives are then either (b) that clusters have more mass than virial estimates based on the cluster galaxy velocities or estimates based on hydrostatic equilibrium [27] of the gas at the measured X-ray temperature (which is surprising since they agree [28]), or (c) that the usual BBN estimate $\Omega_b \approx 0.05(0.5/h)^2$ is wrong. It is interesting that there are indications from weak lensing [29] that at least some clusters may actually have extended halos of dark matter — something that is expected to a greater extent if the dark matter is a mixture of cold and hot components, since the hot component clusters less than the cold [30,31]. If so, the number density of clusters as a function of mass is higher than usually estimated, which has interesting cosmological implications (e.g., $\sigma_8$ is a little higher than usually estimated). It is of course possible that the solution is some combination of alternatives (a), (b), and (c). If none of the alternatives is right, then the only conclusion left is that $\Omega_0 \approx 0.33$. The cluster baryon problem is clearly an issue that deserves very careful examination.

It has recently been argued [32], CHDM models are compatible with the X-ray data within observational uncertainties of both the BBN predictions and X-ray data. Indeed, the rather high baryon fraction $\Omega_b \approx 0.1(0.5/h)^2$ implied by recent measurements of low D/H in two high-redshift Lyman limit systems [33] helps resolve the cluster baryon crisis for $\Omega = 1$ — it is escape route (c) above. With the higher $\Omega_0$ implied by the low D/H, there is now a "baryon cluster crisis", for low $\Omega_0$ models! Even with a baryon fraction at the high end of observations, $f_b \lesssim 0.2(h/0.5)^{-3/2}$, the fair sample hypothesis with this $\Omega_b$ implies $\Omega_0 \gtrsim 0.5(h/0.5)^{-1/2}$.

## 3. $\Lambda$CDM

Although $\Lambda$CDM was considered from the beginning of CDM as a viable alternative to the standard $\Omega = 1$ version [1], the first N-body simulations [34] revealed a problem with the model: the correlation function is too steep and too large at small scales. For example, Efstathiou et al. [4] found that the model with $\Omega_0 = 0.20$ nicely matches the APM angular correlation function $w(\theta)$ on large angular scales ($\theta > 1°$), but disagrees by a factor of 3 with APM results on smaller scales. The disagreement was not considered serious because "the models neglect physics that is likely to be important on small scales where $\xi(r) \gg 1$".

While it is true that one cannot reliably take into account all physics of galaxy formation, that does not mean that we can manipulate the "galaxies" in models without any restrictions. For example, one of the ways out of the problem would be to assume that places with high dark matter density somehow are less efficient in producing galaxies. But Coles [35] showed that if the dark matter density has a Gaussian distribution and the number density of galaxies is any function of local dark matter density ("local bias"), then the correlation function of galaxies cannot be flatter then the correlation function of the dark matter. Coles' results are formally applicable only for a Gaussian density distribution function, which is not the case for real models at a nonlinear scale. But it illustrates our point: just introducing antibias does *not* necessarily save the situation. In the case of a Gaussian distribution it actually makes it worse. Our simulations [5] indicate that Coles' results might be true even for realistic non-Gaussian distributions. This may help explain why we cannot satisfy APM $w(\theta)$ constraints *simultaneously* at large and small scales with any local bias: because the correlation function of dark matter in the $\Lambda$CDM model was already too steep at small scales. (We can reduce small-scale $w(\theta)$, but this will ruin it on large scales.)

Because our numerical and Coles' analytical results were obtained for *local* bias, one might think that in order to reconcile the model with observations we need to appeal to nonlocal effects [36] which might result from photoionization of the ISM by UV photons produced by quasars, AGNs, and young galaxies, or propagation of shock waves produced by multiple supernovae in active galaxies. Actually, neither effect is very efficient in suppressing star formation in high density environments. UV radiation heats gas only to few tens of thousands of degrees. This does affect the formation of small galaxies and delays the time of formation of the first stars in large galaxies, but it cannot change a large galaxy [37,38]. If the gas falls into the gravitational potential of a normal-size galaxy with effective temperature of about $10^6$ K, it does not matter much if it was ionized and preheated. Even if a strong shock is produced by an active galaxy, it is difficult to deliver the shock to a nearby galaxy. Because galaxies are formed in very inhomogeneous environments, a shock produced by one galaxy will have a tendency to damp its energy into a local void, not to propagate into a dense area where another galaxy is forming. Numerical hydro+N-body simulations by Yepes et al. [39], which incorporate effects of UV radiation, star formation, and supernovae explosions, do not show any antibias of luminous matter relative to the dark matter.

In our recent study of ΛCDM [5], we tried to avoid complicated questions about effects of star formation. We used several Particle Mesh dissipationless simulations of various box sizes and resolutions to estimate the nonlinear power spectrum of dark matter, and reinforced the old conclusion that it is not compatible with the observed clustering of galaxies. Our results are shown as the solid line in Figure 2; as expected, the nonlinear spectrum has much extra power on small scales. Thus, this ΛCDM model, which has a formal bias parameter $b \sim 1$, must have extra (anti-) bias. Then we put lower limits on the possible power spectrum of galaxy clustering. This includes two steps:

i) All regions with mass less than $5 \times 10^9 h^{-1} M_\odot$ are assigned zero luminosity because each such region does not have enough mass to produce a galaxy luminous enough to make its way into the CfA or the APM catalogs. It also raises the power spectrum by a factor of four by removing the $\sim 50\%$ of mass in low-density regions.

ii) We assume that in the remaining, higher density regions, the number density of galaxies does not depend on the density of the dark matter.

These purposely unrealistic assumptions suppress the number of galaxies identified in groups and clusters, for example. Because galaxies in high density regions are expected to be more clustered than galaxies in the field, this scheme gives a lower limit on the galaxy clustering predicted by the model. It also gives a significant antibias for galaxies. But even with this antibias, the resulting lower limits on the power spectrum, which are shown in Figure 2 as the dashed curve (from our highest-resolution simulation) and dot-dash curve (from a larger box but lower resolution simulation), are 2–3 times higher than CfA estimates [40] and 3–4 times higher than APM results [3]. (The CfA estimates were obtained by converting the redshift-space power spectrum to real space by comparing redshift-space and real-space power spectra from a large-box low-resolution ΛCDM simulation; both because of the uncertainties inherent in this procedure and because the CfA survey is more shallow, we regard the APM results as being probably more reliable.) A more conventional galaxy identification method (e.g., high peaks) would imply even larger discrepancies.

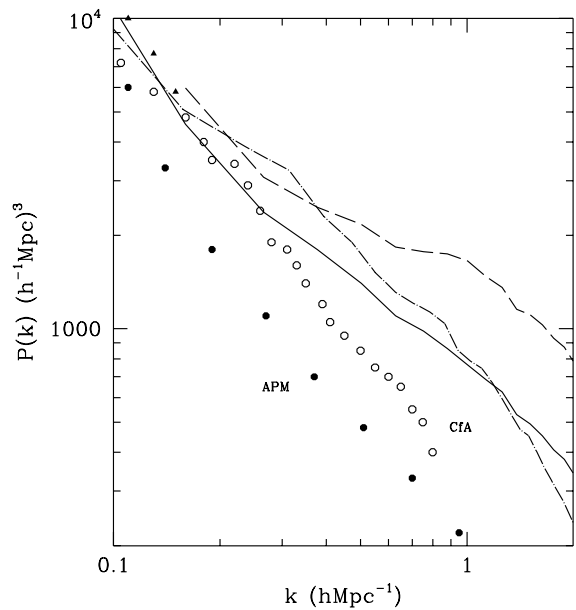

Figure 2. Comparison of the nonlinear power spectrum in ΛCDM model with observational results. Dots are results for the APM galaxy survey. Results for the real-space power spectrum for the CfA survey are shown as open circle ($101h^{-1}$Mpc sample) and triangles ($130h^{-1}$Mpc sample). Formal error bars for each of the surveys are smaller than the difference between the results. The full curve represents the power spectrum of the dark matter. Lower limits on the power spectrum of galaxies predicted by the ΛCDM model are shown as the dashed curve ($\Lambda CDM_f$ simulation in Ref. [5]) and the dot-dashed curve ($\Lambda CDM_c$ simulation).

In our recent paper [5] we show that lower limits on the predicted galaxy correlations in this $\Lambda$CDM model are also much higher than the Stromlo-APM and CfA2 data. It seems that none of the simplest and most attractive schemes for the distribution of galaxies in the model can give correlation functions that agree with observations. Galaxies cannot follow the dark matter. Neither the power spectrum nor the correlation function allow this. The simplest biasing models (halos with overdensity above 200 or density above any reasonable threshold) do not work either: the correlation function discrepancy on 2–3 Mpc scales can be reduced, but the situation on smaller scales gets even worse.

These problems do not mean that all variants of the $\Lambda$CDM model are inconsistent with observations, but it implies that the most attractive variants with large age of the Universe, large Hubble constant, and relatively large cosmological constant are very difficult to reconcile simultaneously with the observed clustering of galaxies and with the number density of galaxy clusters.

## 4. DISCUSSION

Thus, although a $h = 0.7$, $\Omega_0 = 0.3$ $\Lambda$CDM model normalized to COBE nicely fits the data on intermediate scales, including the abundance of rich clusters, the agreement of the linear power spectrum with the data on smaller scales leaves no room for the nonlinear effects that are surely important there. Our minimum N-body estimate of the power spectrum in this model for wavenumbers $k = (0.4 - 1)$ $h^{-1}$ Mpc is at least a factor of 2 too high compared to the CfA results and at least a factor of 3 too high compared to APM. This is true if we assume that the galaxies trace the dark matter without significant bias, but it is also true if we just assume that galaxy density is any monotonically increasing function of dark matter density. It is true even in an extreme model in which we assume that this function is zero for regions of low density and *constant* when the density is above the threshold to produce even a small galaxy. The only way this model can be compatible with the data is for the galaxy formation process to result in strongly scale-dependent anti-biasing on small scales, but this seems hardly compatible with the observations supporting hierarchical scaling and indicating that the number density of luminous galaxies is highest in regions where the dark matter is densest, such as clusters [5]. We suggest that $\Lambda$CDM models will have a better chance of agreeing with observations if they have higher $\Omega_0$, lower $h$, and a tilted spectrum of primordial fluctuations ($n_p \lesssim 0.9$).

It is instructive to compare the $\Omega_0 = 0.3$, $h = 0.7$ $\Lambda$CDM model that we have been considering with standard CDM and with CHDM. At $k = 0.5h$ Mpc$^{-1}$, Figs. 5 and 6 of Ref. [41] show that the $\Omega_\nu = 0.3$ CHDM spectrum and that of a biased CDM model with the same $\sigma_8 = 0.67$ are both in good agreement with the values indicated for the power spectrum $P(k)$ by the APM and CfA data, while the CDM spectrum with $\sigma_8 = 1$ is higher by about a factor of two. As Figure 3 shows, CHDM with $\Omega_\nu = 0.2$ in two neutrino species [12] also gives nonlinear $P(k)$ consistent with the APM data.

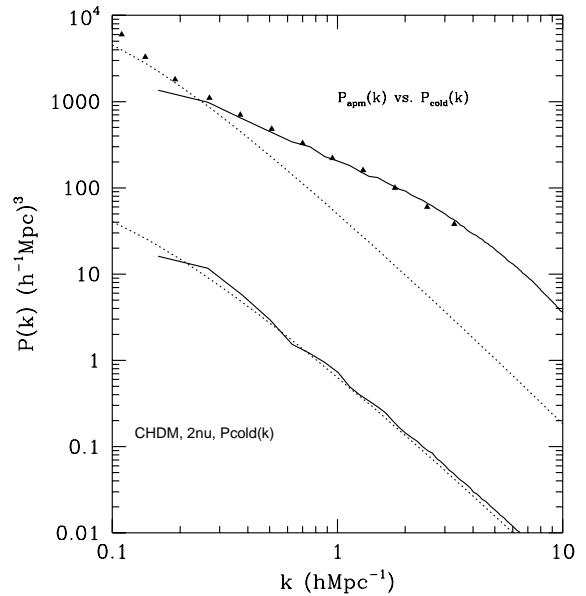

Figure 3. Comparison of APM galaxy power spectrum (triangles) with nonlinear cold particle power spectrum from CHDM model considered in this paper (upper solid curve). The dotted curves are linear theory; upper curves are for $z = 0$, lower curves correspond to the higher redshift $z = 9.9$.

Aside from the Hubble constant, the main potential problem for CHDM appears to be forming enough structure at high redshift. Although as we mentioned above the indications that the amount of gas in damped Lyman $\alpha$ systems is starting to decrease at high redshift $z \gtrsim 3$ seems to be in accord with the available data, the large velocity spread of the associated metal-line systems *may* indicate that these systems are more massive than CHDM would predict (see e.g., [42]). Also, Steidel et al. [43] have found objects by their emitted light at redshifts $z = 3 - 3.5$ apparently with relatively high velocity dispersions (indicated by the equivalent widths of absorption lines), which they tentatively identify as the progenitors of giant elliptical galaxies. *Assuming* that the indicated velocity dispersions are indeed gravitational velocities, Mo & Fukugita (MF) [44] have argued that the abundance of these objects is higher than expected for the COBE-normalized $\Omega = 1$ CDM-type models that can fit the low-redshift data, including CHDM, but in accord with predictions of the $\Lambda$CDM model considered here. (In more detail, the MF analysis disfavors CHDM with $\Omega_\nu \gtrsim 0.2$ in a single species of neutrinos. They apparently would argue that this model is then in difficulty since it overproduces rich clusters — and if that problem were solved with a little tilt $n_p \approx 0.9$, the resulting decrease in fluctuation power on small scales would not lead to formation of enough early objects. However, if $\Omega_\nu \approx 0.2$ is shared between two species of neutrinos, the resulting model appears to be at least marginally consistent with both clusters and the Steidel objects even with the assumptions of MF. The $\Lambda$CDM model with $h = 0.7$ consistent with the most restrictive MF assumptions has $\Omega_0 \gtrsim 0.5$, hence $t_0 \lesssim 12$ Gyr. $\Lambda$CDM models having tilt and lower $h$, and therefore more consistent with the small-scale power constraint discussed above, may also be in trouble with the MF analysis.) But in addition to uncertainties about the actual velocity dispersion and physical size of the Steidel objects, the conclusions of the MF analysis can also be significantly weakened if the gravitational velocities of the observed baryons are systematically higher than the gravitational velocities in the surrounding dark matter halos, as is actually likely to be the case at low redshift for large spiral galaxies [45], and even more so for elliptical galaxies which are largely self-gravitating stellar systems in their central regions.

There is another sort of constraint from observed numbers of high-redshift protogalaxies which disfavors $\Lambda$CDM. The upper limit on the number of $z \gtrsim 4$ objects in the Hubble Deep Field (which presumably correspond to smaller-mass galaxies than most of the Steidel objects) is far lower than the expectations in low-$\Omega_0$ models, especially with a positive cosmological constant, because of the large volume at high redshift in such cosmologies [46]. So evidence from high-redshift objects cuts both ways, and it is too early to tell whether high- or low-$\Omega_0$ models will ultimately be favored. But one sort of constraint on CHDM models is likely to follow both from the high-$z$ data just discussed and from the preliminary indications from the recent results on cosmic microwave anisotropies at and beyond the first acoustic peak [47]: viable CHDM models cannot have much tilt (i.e., $n_p \sim 1$).

Acknowledgments. This work was partially supported by NSF grants at NMSU and UCSC. Simulations were performed on the CM-5 and Convex 3880 at the National Center for Supercomputing Applications, University of Illinois, Champaign-Urbana, IL.

# REFERENCES

1. G.R. Blumenthal, S. Faber, J.R. Primack, and M.J. Rees, Nature 311 (1994) 517.
2. M. Davis, G. Efstathiou, C.S. Frenk, and S.D.M. White, Nature 356 (1992) 489; S. Olivier, J. Primack, G.R. Blumenthal and A. Dekel, Astrophys. J. 408 (1993) 17; S.D.M. White, G. Efstathiou, and C.S. Frenk, Mon. Not. R. Astron. Soc. 262 (1993) 1023.
3. C.M. Baugh and G. Efstathiou, Mon. Not. R. Astron. Soc. 267 (1994) 32.
4. G. Efstathiou, W.J. Sutherland and S.J. Maddox, Nature 348 (1990) 705; L.A. Kofman,

N.Y. Gendin, and N.A. Bahcall, Astrophys. J. 413 (1993) 1; R. Cen, N.Y. Gendin and J.P. Ostriker, Astrophys. J. 417 (1993) 387; R.A. Croft and G. Efstathiou, Mon. Not. R. Astron. Soc. 267 (1994) 390; J.P. Ostriker, and P.J. Steinhardt, Nature 377 (1995) 600; L.M. Krauss, and M.S. Turner, General Relativity and Gravitation 27 (1995) 1137.
5. A. Klypin, J.R. Primack and J. Holtzman, Astrophys. J., in press (July 20, 1996).
6. G. Kauffmann, A. Nusser and M. Steinmetz, astro-ph/9512009.
7. A.R. Liddle, D.H. Lyth, P.T.P. Viana and M. White, astro-ph/9512102.
8. C. Kochanek, astro-ph/9510077) Astrophys. J., in press (1996).
9. J.R. Primack, astro-ph/9604184, to appear in *International School of Physics "Enrico Fermi"*, Course CXXXII: Dark Matter in the Universe, Varenna, eds. S. Bonometto, J.R. Primack, and A. Provenzale (IOS Press, Amsterdam, in press).
10. S.P. Driver, R.A. Windhorst, S. Phillipps, P.D. Bristow, Astrophys. J. 461 (1996) 525.
11. S. Perlmutter, astro-ph/9602122, to appear in *Thermonuclear Supernovae* (NATO ASI), eds. R. Canal, P. Ruiz-LaPuente, and J. Isern; also these proceedings.
12. J.R. Primack, J. Holtzman, A. Klypin, and D.O. Caldwell, Phys. Rev. Lett. 74 (1995) 2160.
13. D. Pogosyan, and A.A. Starobinsky, Astrophys. J. 447, (1995) 465; A. Liddle, D.H. Lyth, R.K. Schaefer, Q. Shafi, and P.T.P Viana, astro-ph/9511057, Mon. Not. R. Astron. Soc., in press, and references therein.
14. S. Ghigna, S. Borgani, S. Bonometto, L. Guzzo, A. Klypin, J.R. Primack, R. Giovanelli and M. Haynes, Astrophys. J. 437 (1994) L71.
15. S. Ghigna, S. Borgani, M. Tucci, S. Bonometto A. Klypin, and J.R. Primack, Mon. Not. R. Astron. Soc., to be submitted (1996).
16. G.M. Fuller, J.R. Primack and Y.-Z. Qian, Phys. Rev. D, 52 (1995) 1288.
17. C. Athanassopoulos et al., Phys. Rev. and Phys. Rev. Lett., submitted (1996) (available at http://nu1.lampf.lanl.gov/lsnd/). Using only the data for $E_\nu = 36 - 60$ MeV, for which the background is lower than the larger neutrino energy range used in last figure in these preprints, lowers the lower limit on $\sin^2 2\theta$ and thus increases the range of allowed $\Delta m^2_{\mu e}$ (private communications from D. Caldwell and S. Yellin, May, (1996).
18. P.J.E. Peebles, *Physical Cosmology* (Princeton University Press, 1993), eq. (5.150).
19. Y. Fukuda, et al., Phys. Lett. B 280 (1994) 146.
20. D. Pogosyan, and A.A. Starobinsky, Astrophys. J. 447 (1995) 465; A. Liddle, et al., Ref. [13].
21. J. Holtzman, Astrophys. J. Suppl. 71 (1989) 1; J. Holtzman, and J.R. Primack, Astrophys. J. 405 (1993) 428; D. Pogosyan, and A.A. Starobinsky, astro-ph/9502019; K.S. Babu, R.K. Schaefer, and Q. Shafi, Phys. Rev. D 53 (1996) 606.
22. L.J. Storrie-Lombardie, R.G. McMahon, M.J. Irwin, and C. Hazard,, Astrophys. J. 427 (1994) L13–16; —, *Proc. ESO Workshop on QSO Absorption Lines*. Newer data suggest that $\Omega_{gas} \approx 3 \times 10^{-3}$ for $z \approx 2 - 3.5$, and either declining or at most remaining constant for higher $z \sim 4-4.5$ (private communications from Lisa Storrie-Lombardi and Art Wolfe, December, 1995 and June, 1996).
23. A. Klypin, S. Borgani, J. Holtzman, and J.R. Primack, Astrophys. J. 444 (1995) 1.
24. S.D.M. White, and C.S. Frenk, Astrophys. J. 379 (1991) 52; S.D.M. White et al., Nature 366 (1993) 429.
25. D.A. White, and A.C. Fabian, Mon. Not. R. Astron. Soc. 273 (1995) 72.
26. K.F. Gunn and P.A. Thomas, astro-ph/9510082.
27. C. Balland and A. Blanchard, astro-ph/9510130.
28. N.A. Bahcall, and L.M. Lubin, Astrophys. J. 426 (1994) 513; M. Bartelmann and R. Narayan, in *Dark Matter*, College Park, MD, October, 1994 (AIP Conference Proceedings 336, 1995) p. 307.
29. G. Fahlman, N. Kaiser, G. Squires and D. Woods, Astrophys. J. 437 (1994) 56;

G. Squires, N. Kaiser, A. Babul, G. Fahlman, D. Woods, M. Neumann and H. Böhringer, Astrophys. J. 461 (1996) 572. Cf. G. Luppino and M. Loenstein contributions in these proceedings.
30. D. Brodbeck, D. Hellinger, R. Nolthenius, J.R. Primack and A. Klypin, Astrophys. J., in press (with accompanying video) (1996).
31. L. Kofman, A. Klypin, D. Pogosyan, and J.P. Henry, astro-ph/9509145.
32. R.W. Strickland and D.N. Schramm, astro-ph/9511111.
33. D. Tytler, X.-M. Fan and S. Burles, Nature, in press (1996); S. Burles, and D. Tytler, astro-ph/9603070, Science, in press.
34. M. Davis, G. Efstathiou, C.S. Frenk, and S.D.M. White, Astrophys. J. 292 (1985) 371.
35. P. Coles, Mon. Not. R. Astron. Soc. 262 (1993) 591.
36. A. Babul, and S.D.M. White, Mon. Not. R. Astron. Soc. 250 (1991) 407; R.G. Bower, P. Coles, C.S. Frenk, and S.D.M. White, Astrophys. J. 405 (1993) 403; S. Ikeuchi and J.P. Ostriker, Astrophys. J. 301 (1986) 522.
37. J. Navarro, and M. Steinmetz, astro-ph/9605043.
38. D.H. Weinberg, L. Hernquist and N. Katz, astro-ph/9604175.
39. G. Yepes, R. Kates, A. Khokhlov and A. Klypin, astro-ph/9605182.
40. C. Park, M.S. Vogeley, M.J. Geller and J.P. Huchra, Astrophys. J. 431 (1994) 569.
41. A. Klypin, R. Nolthenius, and J.R. Primack, astro-ph/9502062, Astrophys. J., in press (1996).
42. L. Lu, W.L.W. Sargent, D.S. Womble, and T.A. Barlow, Astrophys. J. 457 (1996) L1.
43. C.A. Steidel et al., astro-ph/9602024.
44. H.J. Mo and M. Fukugita, Astrophys. J. Lett., in press (1996).
45. J.F. Navarro, C.S. Frenk, and S.D.M. White, Astrophys. J. 462 (1996) 563. Also, D. Zaritsky and S.D.M. White, Astrophys. J. 435 (1994) 599 (and work in preparation) find little correlation between the rotation velocity of optical galaxies and that of their extended dark matter halos. JP thanks M. Steinmetz for emphasizing the relevance of these facts.
46. K.M. Lanzetta, A. Yahil and A. Fernández-Soto, Nature in press (June 27, 1996 issue); A. Yahil, K.M. Lanzetta, A. Campos and A. Fernández-Soto, in preparation (1996).
47. C.B. Netterfield et al., Astrophys. J. 445 (1996) L69. Cf. J. Silk, these proceedings.